One-Dimensional Optical Lattice Clock with a Fermionic $^{171}$Yb Isotope

Takuya Kohno, Masami Yasuda\*, Kazumoto Hosaka, Hajime Inaba, Yoshiaki Nakajima, Feng-Lei Hong

*National Metrology Institute of Japan (NMIJ), National Institute of Advanced Industrial Science and Technology (AIST), Tsukuba Central 3, Ibaraki 305-8563, Japan*

*CREST, Japan Science and Technology Agency, 4-1-8 Honcho Kawaguchi, Saitama 332-0012, Japan*

We demonstrate a one-dimensional optical lattice clock with ultracold $^{171}$Yb atoms, which is free from the linear Zeeman effect. The absolute frequency of the $^1S_0(F = 1/2) - {}^3P_0(F = 1/2)$ clock transition in $^{171}$Yb is determined to be 518 295 836 590 864(28) Hz with respect to the SI second.

\*e-mail address: masami.yasuda@aist.go.jp

The rapid development of research on optical frequency measurement based on femtosecond combs[1,2] has stimulated the field of frequency metrology, and especially research on high-performance optical frequency standards. An optical lattice clock,[3,4] which is expected to have a frequency accuracy of better than $10^{-17}$, has recently attracted great attention. In 2006, a Sr optical lattice clock was recommended by the International Committee for Weights and Measures (CIPM) as one of the secondary representations of the second (candidates for the redefinition of the second), based on measurement results carried out by groups at the Univ. of Tokyo-National Metrology Institute of Japan (NMIJ),[5] JILA (Boulder),[6] and SYRTE (Paris).[7]

Yb has two stable odd isotopes $^{171}$Yb and $^{173}$Yb, which also appear to be excellent candidates for an optical frequency standard with the lattice clock scheme.[8] The absolute frequency of the $^1S_0 - {}^3P_0$ clock transition in $^{171}$Yb and $^{173}$Yb was measured for an atomic cloud released from a magneto-optical trap (MOT) with an uncertainty of 4.4 kHz.[9] On the other hand, an optical lattice clock has also been realized with the even isotope $^{174}$Yb by introducing a static magnetic field to induce a nonzero dipole transition probability for the forbidden clock transition.[10] The absolute frequency of the clock transition of bosonic $^{174}$Yb atoms in an optical lattice has been determined with an uncertainty of 0.8 Hz.[11] The optical lattice induced hyperpolarizability frequency shift uncertainty was measured at less than $7\times10^{-17}$ of the clock frequency (applicable to any Yb isotope),[12] indicating that the Yb optical lattice clock is a promising candidate as a next generation optical frequency standard. Thus far there have been no reports on the absolute frequency of the clock transition of fermionic $^{171}$Yb and $^{173}$Yb atoms in an optical lattice.

In this paper, we report experimental results of the spectroscopy of $^{171}$Yb atoms confined in a one-dimensional (1D) optical lattice and the absolute frequency measurement of the clock transition. We chose $^{171}$Yb because it has a reasonable natural abundance of 14% and a relatively simple $I = 1/2$ spin system, which means we could avoid the need for an extra optical pumping process in the experiment. Figure 1 shows the relevant energy level diagram of $^{171}$Yb. The $^1S_0 - {}^3P_0$ transition has a natural linewidth of 44 mHz and is used as a clock transition. The ultracold atoms for confinement in an optical lattice were prepared by using two stages of MOT. Firstly, by using the strong dipole-allowed transition ($^1S_0 - {}^1P_1$; 399 nm, natural linewidth of 28 MHz), we decelerated Yb atoms in an atomic beam (from 770 K to 1 K) with a Zeeman slower. The atoms were then loaded into the 1$^{st}$ stage MOT using the same transition. We used a 399-nm external cavity diode laser stabilized with an Yb hollow cathode lamp for the Zeeman slower and the 1$^{st}$ stage MOT.[13] After switching the laser light and the magnetic field gradient, the atoms were cooled further in the 2$^{nd}$ stage MOT using the spin-forbidden transition ($^1S_0 - {}^3P_1$; 556 nm, natural linewidth of 182 kHz). To obtain the 2$^{nd}$ stage MOT, it was also necessary to add a compensation dc magnetic field to the quadruple magnetic field. The green light source at 556 nm was a frequency-doubled fiber DFB laser, which was phase-locked to a home-made optical frequency comb using a mode-locked fiber laser (fiber comb)[14] for frequency stabilization. The Yb atoms were cooled down to 6.5 mK and 40 μK, in the 1$^{st}$ and 2$^{nd}$ stage MOT, respectively.

Figure 2 is a schematic diagram of our experimental setup for spectroscopy and frequency measurement. The clock laser at 578 nm was obtained by using a sum-frequency generation (SFG) scheme with a periodically-poled-lithium-niobate (PPLN) waveguide (WG) device.[15] We used a 1030-nm Yb:YAG laser and a 1319-nm

Nd:YAG laser to generate the SFG light.  The SFG light was frequency stabilized to a high-finesse cavity made from ultra-low expansion (ULE) glass to reduce the laser linewidth.  The clock laser frequency was measured with another fiber comb.  Part of the broadened fiber comb at 1156 nm was frequency-doubled to 578 nm for the measurement.  The fiber comb was phase-locked to coordinated universal time [UTC (NMIJ)], which was linked to international atomic time (TAI).

The clock laser light was transferred using a 44-m-long single-mode fiber with fiber noise cancellation, and steered into the Yb optical lattice through an acousto-optic modulator (AOM) for both frequency tuning and intensity switching.  The lattice laser at 759 nm was a cw Ti:sapphire laser, and it was focused into the vacuum chamber (power ~ 400 mW, radius = 25 μm) together with the clock laser.  Only the lattice laser was reflected back to the chamber by using a curved dichroic mirror to form the optical lattice.

A 399-nm probe laser was used to detect the spectrum with an approach based on the electron shelving technique.  After the cold atoms were loaded into the lattice, the AOM was turned on to apply a 578-nm light pulse (width = 100 ms, intensity ~ 2.3 mW/cm$^2$) to the atoms.  The frequency of the Rabi oscillations was estimated to be ~ 1 kHz.  This pulse excites the atoms to the $^3P_0$ state when the clock laser is on resonance. We then used the 399-nm probe light to excite the atoms remaining in the ground state to the $^1P_1$ state to allow us to observe the fluorescence using a photo-multiplier tube. Therefore, the intensity of the fluorescence was proportional to the ground state population.  Since the atoms excited to the $^3P_0$ state will remain for much longer than the measurement cycle, the ground state population and hence the 399-nm fluorescence will decrease when the clock laser is on resonance.  Figure 3 shows the spectrum of

$^{171}$Yb observed by using the electron shelving technique.   Each data point is an average of 10 measurements, where each measurement required a total of 1.1 s for cooling, capture and detection.   The laser frequency was tuned with a step of 80 Hz by scanning the AOM frequency.   The drift of the ULE frequency was compensated with the comb measurement data.   The four components of the spectrum correspond to transitions between the nuclear Zeeman sublevels of $^{171}$Yb in a magnetic field.   The magnetic field was a residual dc field, containing both the compensation dc magnetic field and a stray magnetic field, when the quadruple field of the 2$^{nd}$ stage MOT has been turned off.   The center frequency of each component was obtained by fitting the experimental data to a Lorentzian function.   The frequency of the clock transition was calculated by using the average of the frequencies of all four components to cancel the first order Zeeman shifts.[5]

To determine the magic wavelength, we measured the light shift as a function of the lattice laser intensity at several wavelengths near 759.3 nm, as shown in the inset of Fig. 4.   Each data point was calculated from one measured spectrum (similar to the spectrum in Fig. 3), where the uncertainty came from the center fitting of the Zeeman components.   The slope of the fitted lines in the inset, which is the differential light shift, moved from positive, through zero at the magic wavelength, to negative, when the lattice wavelength was decreased.   The differential light shift was then fitted to determine the magic wavelength (Fig. 4).   The determined magic wavelength was 759.353(3) nm.   We note that the magic wavelength uncertainty introduced a scalar light shift uncertainty of 14 Hz to the clock frequency (see Table I).

Figure 5 shows the absolute frequency of 12 measurements, where each data point was calculated from one measured spectrum.   When we average the 12 measurements we obtain an average frequency of 518 295 836 590 864 Hz. The standard deviation of the mean is 5.6 Hz.

The corrections and uncertainties for the $^{171}$Yb lattice clock are listed in Table I. The calculated blackbody radiation shift was -1.32 Hz[16] for the temperature of the vacuum chamber ($T$ = 298.9 K) in which the optical lattice clock was operated. The uncertainty is mainly contributed by the theoretical calculation[16] when the uncertainty of the temperature estimation is less than 3 K. For the gravitational shift, the correction was calculated using the height of the optical lattice from the geoid surface (19.3 m) while the uncertainty was estimated from that of the height measurement (0.5 m). The second order Zeeman shift was estimated to be - 0.4 Hz using a magnetic field of 2.5 G, which was calculated from the Zeeman splitting. The light shift induced by the clock laser was estimated to be + 40 mHz using a typical clock laser intensity of 2.3 mW/cm$^2$ . In the determination of the clock transition frequency using the observed spectrum (the paper lock), the data has a frequency step of 80 Hz as shown in Fig. 3. The paper lock error was calculated using the uncertainty of the uniform distribution (80/2√3 Hz). The uncertainty of the UTC (NMIJ) was less than $1\times10^{-14}$ during the measurement, which led to an uncertainty of < 5 Hz in the Yb absolute frequency. The collision shift was not observed when we changed the atom number loaded into the lattice, and will be eliminated by introducing the spin-polarized 1D lattice scheme[5] in the next step. The vector light shift has been estimated to be below the mHz level.[8] The second order Doppler shift was estimated to be $< 1\times10^{-20}$.

We obtained a total systematic uncertainty of 27 Hz in Table I. This uncertainty was then combined with the statistical uncertainty of 5.6 Hz obtained from Fig. 5, and gave a final combined uncertainty of 28 Hz, which is consistent with the MOT frequency measurement result[9] within their measurement uncertainty of 4.4 kHz. The present measurement uncertaitny (28 Hz, fractionally $5.4\times10^{-14}$) is expected to be reduced to the Cs limit by performing feedback control of the clock frequency using the AOM and improving the signal-to-noise ratio of the spectrum. Finally, the determined absolute frequency of the $^1S_0(F = 1/2) - {}^3P_0(F = 1/2)$ clock transition in $^{171}$Yb is 518 295 836 590 864(28) Hz. The measurement results will be reported to the CIPM for a discussion of new frequency standards and should have a great impact on Yb lattice clock research.

Acknowledgments  We are grateful to M. Imae, Y. Fujii and T. Suzuyama for their work in maintaining UTC (NMIJ). We thank A. Onae, S. Ohshima and D. Akamatsu for helpful discussions. We also thank Y. Kawakami and S. Nagahama for supplying the 399-nm laser diodes.

3

Figure captions

Fig. 1. Energy levels of $^{171}$Yb. Wavelengths and natural linewidths are indicated for the relevant cooling, trapping, and clock transitions.

Fig. 2. Schematic diagram of the experimental setup.  SFG, sum-frequency generation; PPLN, periodically poled lithium niobate; WG, waveguide; ULE, ultra-low expansion; SM, single mode; AOM, acousto-optic modulator; PMT, photo-multiplier tube; UTC, coordinated universal time; NMIJ, National Metrology Institute of Japan.

Fig. 3. Ground state population as a function of clock laser frequency $f$.  The four components in the spectrum correspond to transitions between nuclear Zeeman sublevels of $^{171}$Yb in a magnetic field.

Fig. 4. The light shift of the clock transition measured as a function of the lattice laser intensity at several wavelengths near 759.3 nm.  The magic wavelength is determined as 759.353(3) nm.

Fig. 5. Absolute frequency measurement of the $^1S_0(F = 1/2) - {}^3P_0(F = 1/2)$ transition in $^{171}$Yb.  The average of the 12 measurements is 518 295 836 590 864 Hz.

Table I. Systematic corrections and uncertainties for the $^{171}$Yb lattice clock.

| Effect | Correction (Hz) | Uncertainty (Hz) |
|---|---|---|
| Blackbody radiation shift | + 1.32 | 0.13 |
| Gravitational shift | − 1.19 | 0.03 |
| 2nd order Zeeman shift | + 0.4 | 0.05 |
| Scalar light shift | 0 | 14 |
| Clock laser light shift | − 0.04 | < 0.01 |
| Paper lock error | 0 | 23 |
| UTC (NMIJ) | 0 | 5 |
| Total | + 0.49 | 27 |

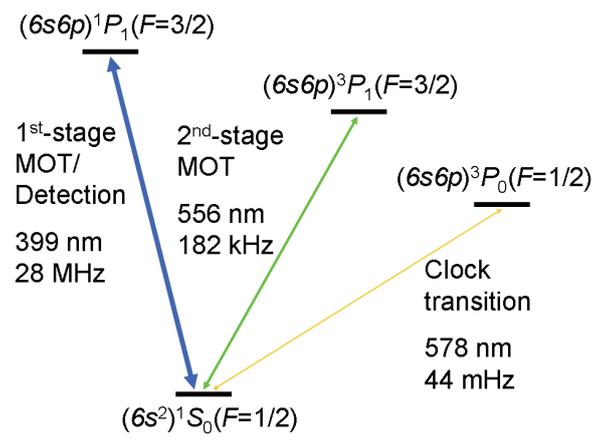

Fig. 1

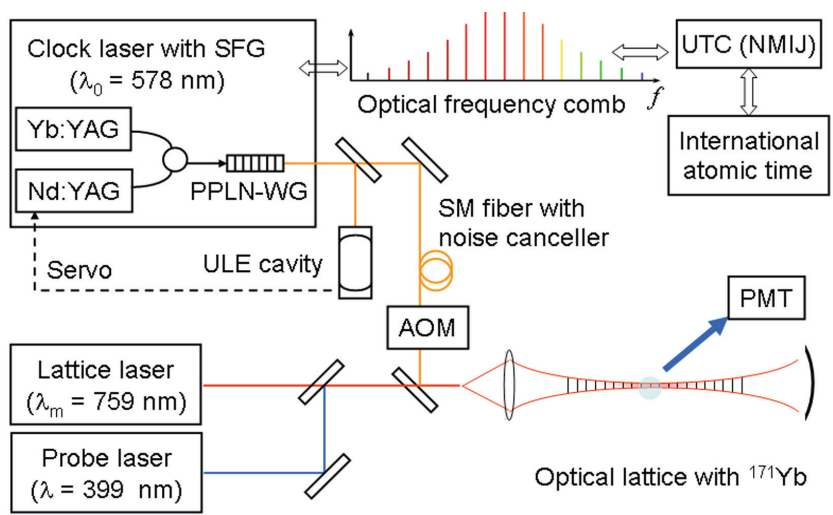

Fig. 2

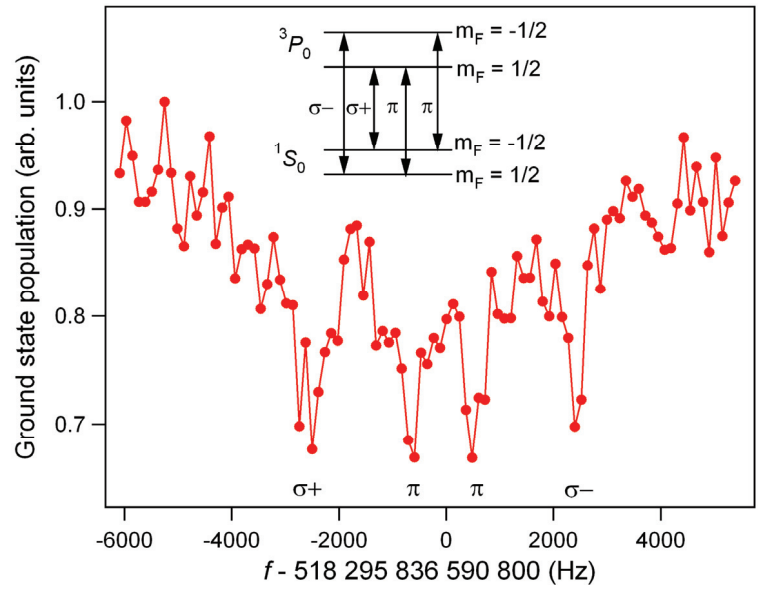

Fig. 3

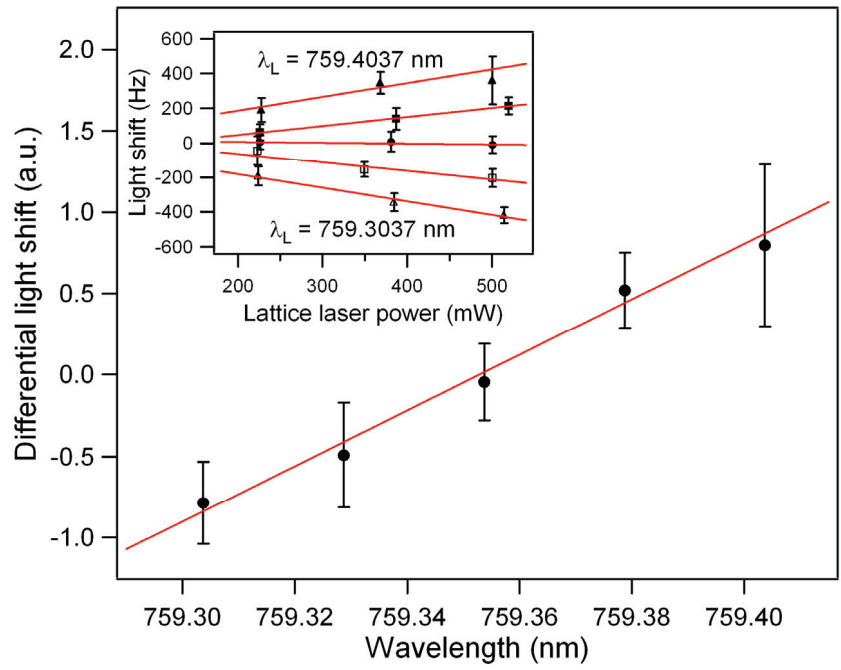

Fig. 4

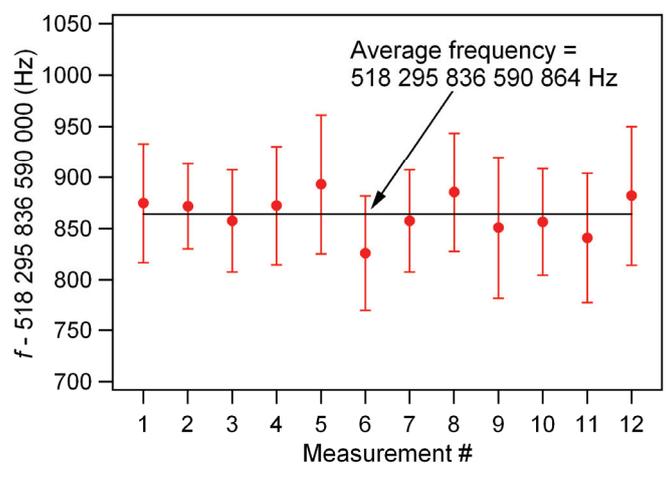

Fig. 5